\documentclass[reprint,aps,pra,amsmath,amssymb,showpacs,superscriptaddress,]{revtex4-1}
\usepackage{graphicx}
\usepackage{dcolumn}
\usepackage{bm}
\usepackage{color}
\usepackage{upgreek}

\begin{document}

\title{High harmonic generation from periodic potentials driven by few-cycle laser pulses}

\author{Zhong Guan}\affiliation{State Key Laboratory of Magnetic Resonance and Atomic and Molecular Physics, Wuhan Institute of Physics and Mathematics, Chinese Academy of Sciences, Wuhan 430071, Peoples Republic of China}\affiliation{College of Physics and Electronic engineering, Northwest Normal University, Lanzhou 730070, Peoples Republic of China}

\author{Xiao-Xin Zhou}\affiliation{College of Physics and Electronic engineering, Northwest Normal University, Lanzhou 730070, Peoples Republic of China}

\author{Xue-Bin Bian}\email{xuebin.bian@wipm.ac.cn}\affiliation{State Key Laboratory of Magnetic Resonance and Atomic and Molecular Physics, Wuhan Institute of Physics and Mathematics, Chinese Academy of Sciences, Wuhan 430071, Peoples Republic of China}

\begin{abstract}

We investigate the high harmonic generation (HHG) from solids by simulating the dynamics of a single active electron in periodic potentials. The corresponding time-dependent Schr\"odinger equations (TDSE) are solved numerically by using B-spline basis sets in coordinate space. The energy band structure and wave vectors can be directly retrived from the eigenfunctions. The harmonic spectra obtained agree well with the results simulated by TDSE in $k$ space using Bloch states and show a two-plateau structure. Both of the cutoff energies of the two plateaus in the harmonic spectrum scale linearly with the field strength. We also study HHG driven by intense few-cycle laser pulses and find that the cutoff energy of the harmonic spectrum is as sensitive to the changes of the carrier envelope phase, as to HHG from gas samples, which suggests recollision pictures in HHG as found by recent experiments (Nature {\bf 522}, 462 (2015)).


\end{abstract}

\maketitle

\section{INTRODUCTION}\label{I}

The process of high harmonic generation (HHG) has been studied extensively \cite{Corkum0, Krauz} in the past several decades, which has become one of the most important research areas in ultrafast atomic and molecular physics. HHG has been a tabletop coherent x-ray source. It can be used to generate attosecond (1 as=10$^{-18}$ s) laser pulses \cite{Sansone}, image molecular structures \cite{Itatani, Bian4}, and so on. Whereas, because of the high nonlinearity of the HHG process \cite{Corkum1}, the intensity of generated harmonics is still too low, which limits the applications of HHG.

Recently, experiments \cite{Ghimire} showed that HHG can be generated from bulk crystals. Due to high density of solid materials, it is possible to produce HHG with higher efficiency. In addition, by analysis of the spectra of HHG, it may be possible to study internal structures of solid materials.

HHG in solids is generally considered to involve two contributions: interband and  intraband currents. Recent experiments \cite{Vampa} revealed that electron-hole recollision leads to the interband current, while intraband current is thought as a result of Bloch oscillations \cite{Vampa1, Vampa2} in the same band. Both processes are illustrated in Fig. \ref{Fig1}. Ghimire \textit{et al}. \cite{Ghimire2} suggested that laser-driven Bloch oscillations of an electron wave packet on a single conduction lead to current. Wu \textit{et al}. \cite{Wu} suggested that a primary plateau is due to the coupling of the valence band to the first conduction band and a weaker second plateau is due to coupling to higher-lying conduction bands. 2-band and 3-band models are calculated by McDonald \textit{et al}. \cite{McDonald}. Their results indicate that the first plateau arises from electron-hole recollision, while the higher plateaus arise from interband Bloch oscillations.

Until now, there are two main theoretical models: the many-electron model and the single-electron model. The former is based on the semiconductor Bloch equation (SBE) \cite{Lindberg}, which has been successfully applied in solids and semiconductor physics, such as semiconductor excitons, many-body correlations \cite{Klimov}. However, it has limitations. Due to the interactions between electrons in solids, it is difficult to obtain accurate solutions to many-electron systems. On the other hand, it may take a long time to solve the related problems. The single-electron model exhibits its advantages in addressing electron dynamics in solids. It is based on energy-band theory, which treats motions of each electron in solids independently in an effective potential. For ideal crystals, atoms are arranged in a regular rule, which shows periodicity. As long as the effective potential is determined, the related problems can be solved directly.

In this work, we use B-spline basis functions to solve the single-electron time-dependent Schr\"odinger equations (TDSE) in coordinate space by using periodic potentials to simulate HHG processes in solids. Inter- and intra- band transitions, cutoff dependence, and carrier envelope phase (CEP) effects in short laser pulses are studied.

The article is organized as follows. In Sec. \ref{II}, we review the theoretical framework and exhibit results by comparing different methods. In Sec. \ref{III}, we show numerical simulations for HHG in short pulses. We conclude our work in Sec. \ref{IV}.

\section{Theoretical methods}\label{II}

\begin{figure}
\centering\includegraphics[width=9 cm,height=9 cm]{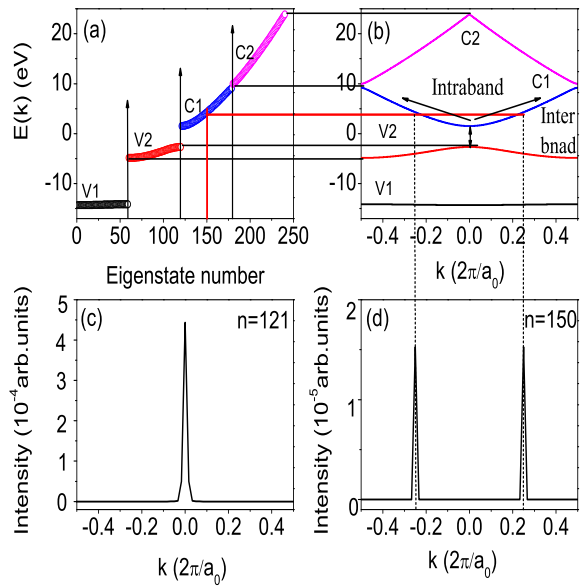}
\caption{(Color online) Band structure calculated by our proposed method compared to results obtained by Bloch states expansion. (a) Four bands calculated by using B-spline basis. (b) Four bands calculated by using Bloch state basis. Interband dynamics describes the transition of electrons between different bands, while intraband transitions involve dynamics of electrons in the same band. (c) Fast Fourier transformation (FFT) of the initial state in our calculations. (d) FFT of the 150th eigenstate.} \label{Fig1}
\end{figure}

Our theoretical method is based on the single active electron model, we describe the laser-crystal interaction in one-dimensional systems with laser polarization in the crystal plane. In the length gauge, the time dependent Hamiltonian is written as
\begin{equation}\label{E1}
\hat{H}(t)=\hat{H}_0+ e\cdot xF(t),
\end{equation}
where $\hat{H}_0=\frac{{\hat{p}^2}}{2m}+V(x)$, and $V(x)$ is a periodic potential of the lattice. In our calculations, we choose the Mathieu-type potential \cite{Wu}. The specific form is $V(x)=-V_0[1+\cos(2\pi x/a_0)]$, with $V_0=0.37$ a.u., and lattice constant $a_0=8$ a.u.

In absence of the external field, the time-independent Schr\"odinger equation can be written as

\begin{equation}\label{E2}
H_0\phi_{n}(x)=E_{n}\phi_{n}(x).
\end{equation}

We use B-spline functions \cite{Bachau} to expand the time-independent wave function,

\begin{equation}\label{E3}
\phi_{n}(x)=\sum_{i=1}^{N_{max}}c_{i}B_{i}(x).
\end{equation}

Substituting Eq.(\ref{E3}) into Eq.(\ref{E2}), we obtain matrix equation

\begin{equation}
HC=ESC,
\end{equation}

where $C$ is the column matrix, $H$ and $S$ are $N\times N$ square matrix, respectively. The matrix elements are
\begin{eqnarray}
H_{ji}& = &\int B_{j}(x)\left[-\frac{d^2{}}{dx^2}-V(x)\right]B_{i}(x)dx, \\
S_{ji}& = &\int B_{j}(x)B_{i}(x)dx.
\end{eqnarray}

We use 3600 B-splines to calculate its eigenvalues in the space region [-240, 240] a.u. and obtain the energy bands, which are illustrated in Fig. \ref{Fig1}. As we know the Bloch-state basis is widely used in periodic systems and can be expanded in plane waves. All calculations based on it can be performed in wave vector $k$ space, and the integral of currents involves the whole first Brillouin zone. In contrast, we use B-spline basis to perform all calculations in coordinate space, in which the periodic property is not explicit in basis. However, calculations in coordinate space and wave vector space are equivalent. To demonstrate it, we also plot the energy bands obtained by the Bloch-state expansion in $k$ space, comparing the results obtained by the B-spline method described above. One may find that the number of bands, the width of the bands and band gaps agree well with each other. In our coordinate space method, the wave vector is not explicit. However, it can be easily extracted by Fast Fourier transformation (FFT) of the eigenfunction. We choose arbitrarily an eigenfunction and transform it from coordinate space to wave vector space by FFT. One can see in Fig. \ref{Fig1}(d) that one eigenstate corresponds to two wave vectors, which agrees well with the Bloch-state expansion \cite{Wu}.

In the laser fields, electrons oscillating in the valence band have probabilities to tunnel to conduction bands. Because tunnelling probabilities are related to the energy gap, only a small portion populated near $k$=0 in band 2 can tunnel from valence to conduction bands with the laser parameters used in the current work. So we choose an initial state calculated by our B-spline method with $k$=0 in band 2 with the smallest band gap . Its FFT is illustrated in Fig. \ref{Fig1}(c).

We use Crank-Nicolson (CN) method \cite{Bian2} to calculate the time-dependent coefficients $C(t)$.

\begin{equation}
C(t+\triangle t)=\frac{S-iH(t+\triangle t/2)\triangle t/2}{S+iH(t+\triangle t/2)\triangle t/2}C(t).
\end{equation}

After obtaining the coefficients at an arbitrary time, we calculate the time-dependent laser-induced currents
\begin{equation}\label{E4}
j(t)=-\frac{e}{m}\{\left[\langle \psi{}(t)|\hat{p}|\psi{}(t)\rangle\right]\}.
\end{equation}

As we know there are many bands in solid crystals, we have to extract each band's information from the eigensystem. Each band group can be distinguished easily as illustrated in Fig. \ref{Fig1}. For example, band 1 corresponds to the state number 0$\sim$59, band 2 corresponds to state number 60$\sim$121. The intraband contribution to the current involves transitions between states in the same band, while the interband contribution involves transitions between states in different bands.

We can use the eigenstates to expand the time-dependent wave function
\begin{equation}\label{E5}
|\psi{}(t)\rangle=\sum_{b}\sum_{n}a_{n}^{b}(t) |{\phi}_{n}^{b}{}\rangle,
\end{equation}
where $b$ stands for the band number.

By substituting Eq.(\ref{E5}) into Eq.(\ref{E4}), the total current is written as
\begin{equation}
j(t)=-\frac{e}{m}\sum_{bb'}\sum_{nn'}\langle \psi{}(t)|\phi_{n}^{b}{}\rangle\langle \phi_{n}^{b}{}|\hat{p}|\phi_{n'}^{b'}{}\rangle\langle\phi_{n'}^{b'}|\psi{}(t)\rangle.
\end{equation}

Each band may correspond to many eigenstates, so the current of interband involving transitions between different states in different bands can be written as
\begin{equation}
j_{\textrm{inter}}= -\frac{e}{m}\sum_{bb'\atop b\neq b'}\sum_{nn'}\langle \psi{}(t)|\phi_{n}^{b}{}\rangle\langle \phi_{n}^{b}{}|\hat{p}|\phi_{n'}^{b'}{}\rangle\langle\phi_{n'}^{b'}|\psi{}(t)\rangle.
\end{equation}

The current of intraband involving transitions between states in the same band can be written as
\begin{equation}
j_{\textrm{intra}}= -\frac{e}{m}\sum_{b}\sum_{nn'}\langle \psi{}(t)|\phi_{n}^{b}{}\rangle\langle \phi_{n}^{b}{}|\hat{p}|\phi_{n'}^{b}{}\rangle\langle\phi_{n'}^{b}|\psi{}(t)\rangle.
\end{equation}

\section{NUMERICAL SIMULATIONS}\label{III}
\begin{figure}

\centering\includegraphics[width=9 cm,height=5 cm]{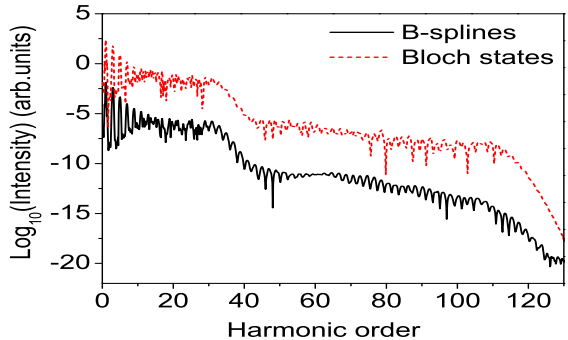}
\caption{(Color online) High-harmonic spectra of laser-induced currents. The laser intensity $I=8.087\times 10^{11}$ W/cm$^2$, wavelength $\lambda=3.2$ $\upmu$m. The upper red dashed line shows harmonic spectrum calculated by Bloch-state basis, while the lower solid black line is the harmonic spectrum calculated by using B-spline basis. For clarity, the latter is down shifted.}
 \label{Fig2}
\end{figure}
\begin{figure}
\centering\includegraphics[width=8.5 cm,height=8.5 cm]{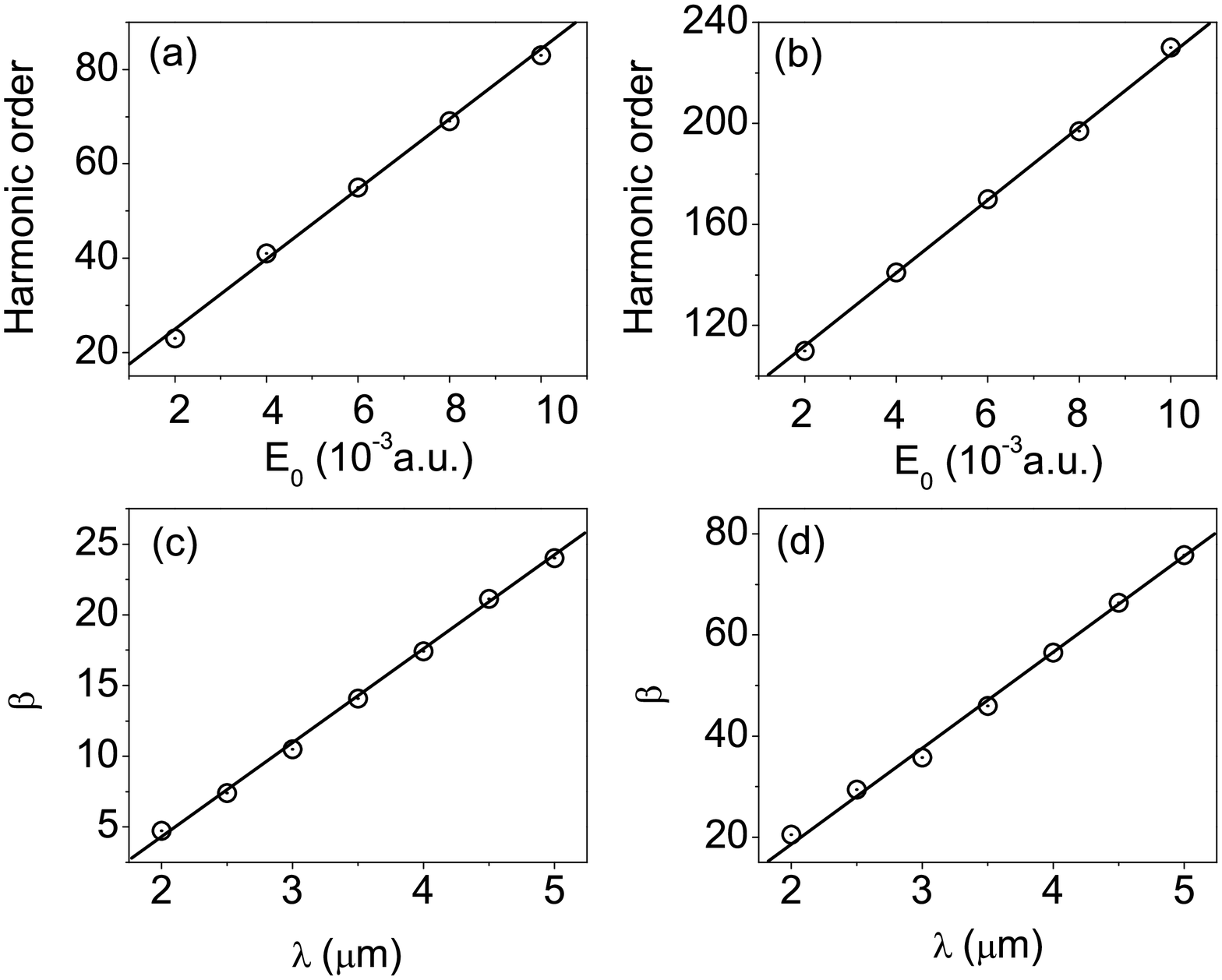}
\caption{ (a) Intensity dependence of the cutoff of the first plateau. (b) Intensity dependence of the cutoff of the second plateau. (c) Scaling coefficient of the cutoff of the first plateau as a function of the wavelengths. (d) Scaling coefficient of the cutoff of the second plateau as a function of the wavelengths.  $\eta_{\textrm{cutoff}}\simeq\beta\hbar\omega_{B}$, $\omega_{B}$ is the Bloch frequency.} \label{Fig3}
\end{figure}

\begin{figure}
\centering\includegraphics[width=9 cm,height=6.5 cm]{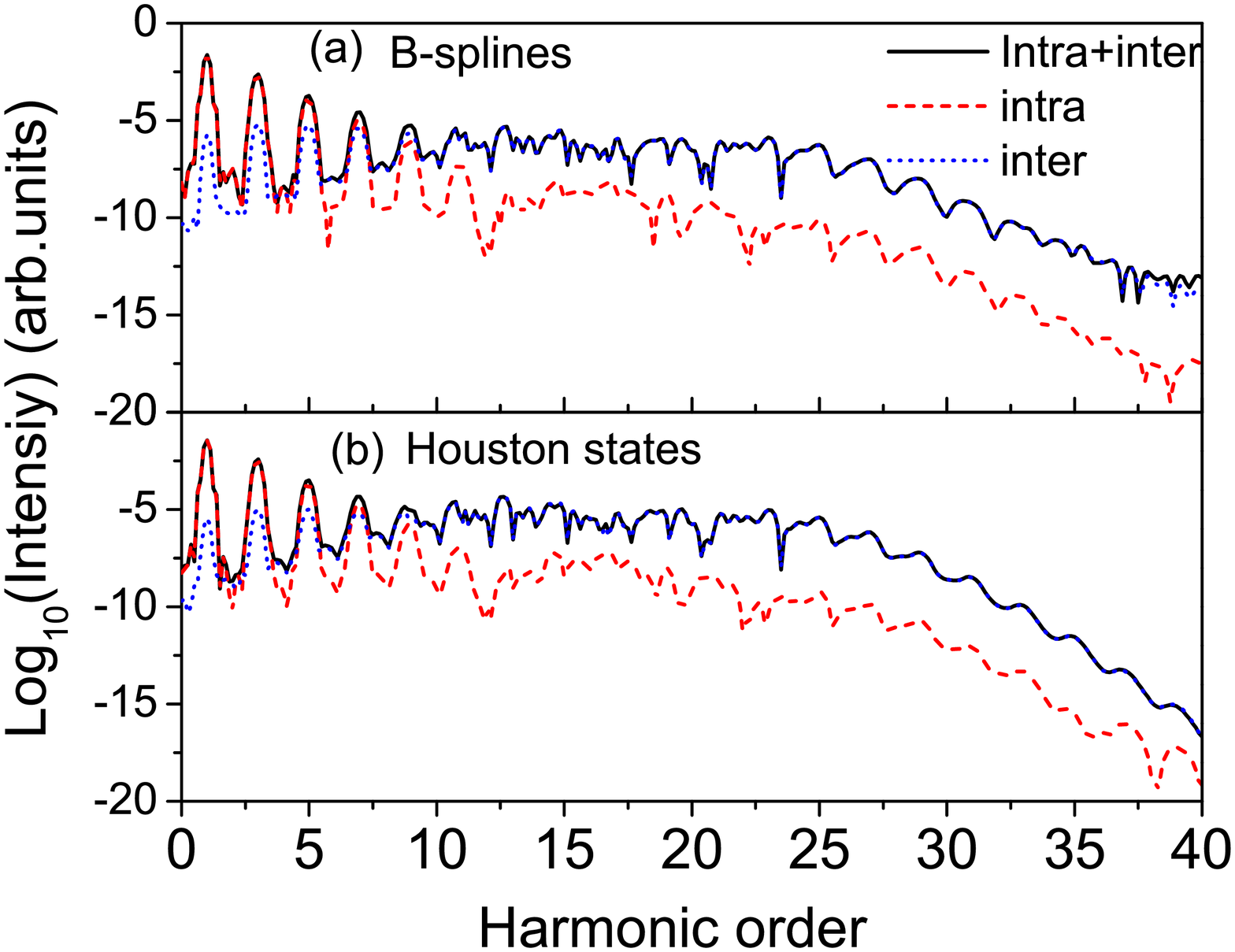}
\caption{(Color online) (a) HHG induced by intraband current and interband current in B-spline method. (b) HHG induced by intraband currents and interband currents in Houston basis. The laser parameters are the same as those used in Fig. \ref{Fig2}.} \label{Fig4}
\end{figure}

In this paper, the laser pulse we use has a $\cos^2$ envelope with durations from 2 cycles to 12 cycles. In our calculations, the absorbing function that we use is a $\cos^{1/8}$ function with $|x|>$200 a.u. to reduce artificial reflections from the boundary. We have considered laser intensities between $1.404\times 10^{11}$ W/cm$^2$ and $3.51\times 10^{12}$ W/cm$^2$, and laser wavelengths between 2 $\upmu$m and 5 $\upmu$m. Before calculating harmonic spectrum, we multiply $j(t)$ by a Hanning window.

\subsection{Comparison to Bloch and Houston state expansions}

To check the validity of our proposed method above, we used different methods to do comparisons. The laser intensity is $I=8.087\times 10^{11}$ W/cm$^2$, and the laser wavelength $\lambda = 3.2 \upmu$m. The total pulse duration is 8 cycles.

The HHG spectrum obtained by our B-spline method is presented in Fig. \ref{Fig2}. One may find a two-plateau structure clearly.  The intensity of the first plateau is around 5 orders higher than the second plateau. This may be the reason why the second plateau has never been observed experimentally. We also show the HHG spectrum calculated by Bloch states introduced by Wu \textit{et al.} \cite{Wu}, which is also illustrated in Fig. \ref{Fig2}. The cutoff energies of the two plateaus, the overall trend and some details agree well with each other, showing the validity of the current method. The first plateau is mainly contributed by the interband transition of bands 2 and 3. The second plateau is mainly generated from higher band transitions. For details we refer readers to Ref. \cite{Wu}.

We also calculated the cutoff energy dependence of the first and second plateaus. We fix the laser wavelength at $\lambda = 4$ $\upmu$m  and change laser intensity from $1.404\times 10^{11}$ W/cm$^2$ to $3.51\times 10^{12}$ W/cm$^2$. Fig. \ref{Fig3}(a) shows that the cutoff of the first plateau increases linearly with field strength. The result is consistent with experiments \cite{Ghimire}. For the second plateau, it is also linearly dependent on field strength, but the ratio is different from the first one. We also investigate the wavelength dependence of the cutoff energy of the first and the second plateaus. The laser intensity is fixed at $I = 8.087\times 10^{11} W/cm^2$, and we change laser wavelengths from 2 $\upmu$m to 5 $\upmu$m. we find the cutoff energy of two plateaus also linearly depend on wavelengths. The results are shown in Figs. \ref{Fig3}(c) and \ref{Fig3}(d). The cutoff energy is related to the Bloch frequency $\omega_{B}$, $\eta_{\textrm{cutoff}}\approx \beta\hbar\omega_{B}$.

The results of this work are consistent with Ref. \cite{Wu}. The cutoff energies of the two plateaus depend linearly on both the wavelength and strength:

\begin{equation}
\eta_{\textrm{cutoff}}\propto (\lambda E)
\end{equation}

Although Bloch-state basis is computationally convenient, it can not be used to separate the intraband and interband currents. As described in the above section, we present the intraband and interband contributions to HHG in the first plateau by B-spline method in Fig. \ref{Fig4}(a). The laser parameters are the same as those in Fig. \ref{Fig2}. Intraband transitions play key roles in the first few harmonics, while interband transitions are dominant in the higher order harmonics. We also show the results by Houston basis \cite{Wu} in Fig. \ref{Fig4} (b). One can see that the results by the different methods agree well. However, the Houston-state treatment becomes numerically unstable as the laser intensity increases \cite{Wu}. B-spline method can overcome this disadvantage, which is a general method successfully used even in non-periodic systems \cite{Bachau}.

\subsection{CEP dependence of HHG}
CEP effects in ultrashort laser pulses have been well studied in HHG in gas phase. To our knowledge, the CEP effects on HHG from solid phase have not been investigated.

\begin{figure}

\centering\includegraphics[width=9 cm,height=4 cm]{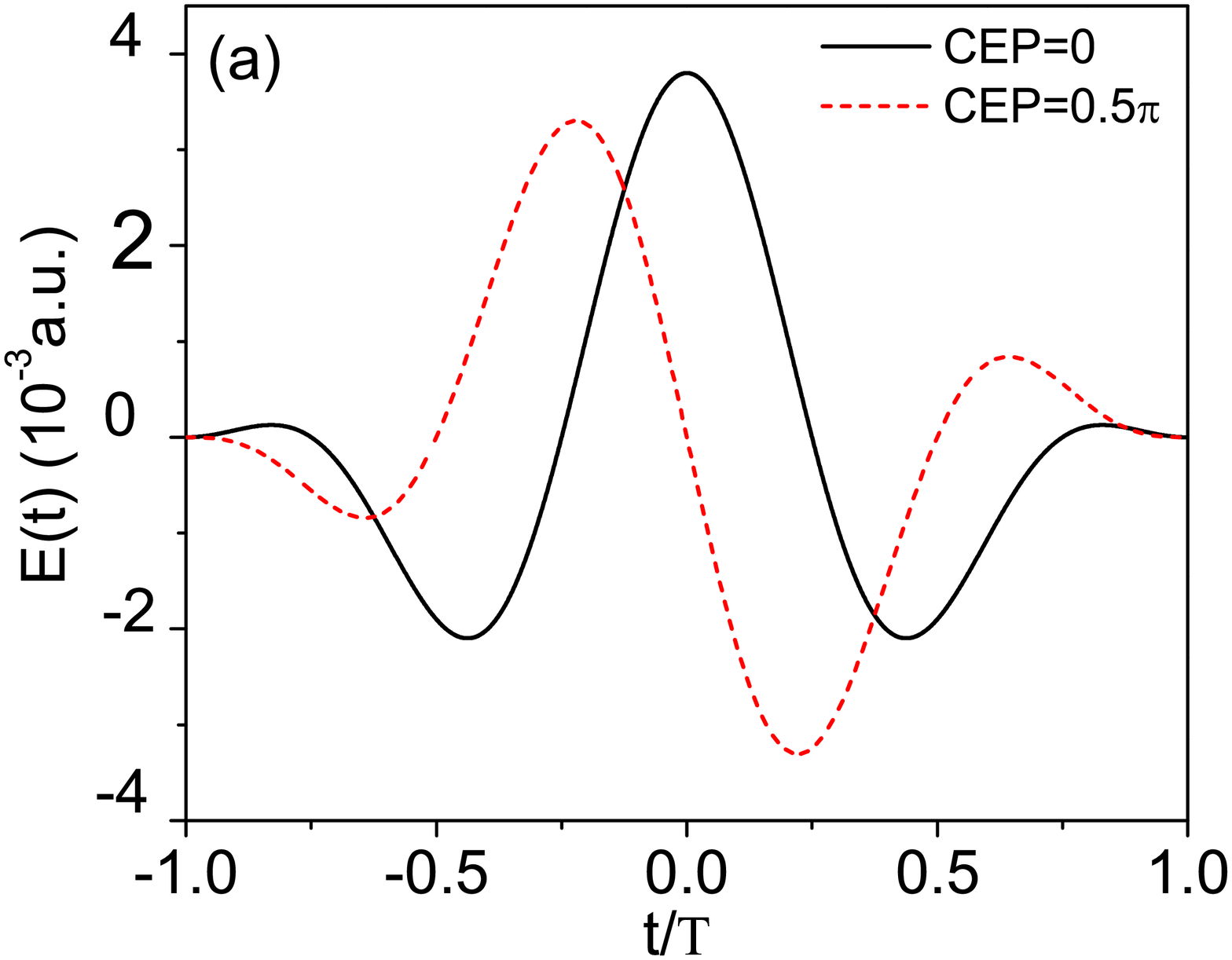}
\centering\includegraphics[width=9 cm,height=4 cm]{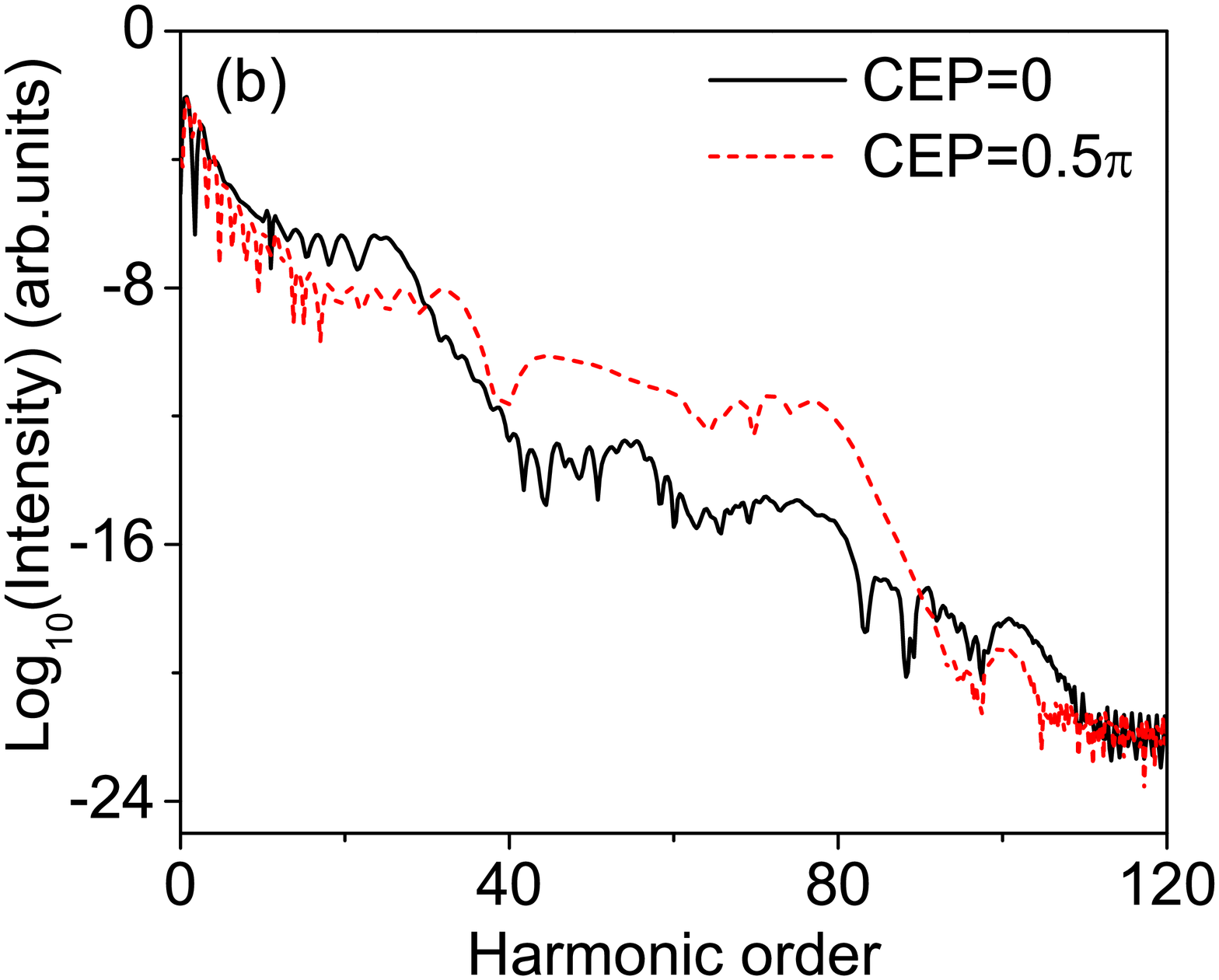}
\caption{(Color online) CEP effect on HHG from periodic potentials. The laser intensity $I=5.068\times 10^{11}$ W/cm$^2$ and laser wavelength $\lambda$=3.2 $\upmu$m. The total pulse duration is 2 cycles. (a) Laser pulses with different CEPs. (b) HHG obtained with different CEPs.} \label{Fig5}
\end{figure}

In our calculations, we use laser intensity $I=5.068\times 10^{11}$ W/cm$^2$ and laser wavelength $\lambda$=3.2 $\upmu$m. The total pulse duration is 2 cycles. The electric fields with the CEPs $\phi=0$ and $0.5\pi$ are presented in Fig. \ref{Fig5}(a). The corresponding HHG spectra are illustrated in Fig. \ref{Fig5}(b). One can see that the cutoff and intensity of HHG are very sensitive to CEPs. Especially for the second plateau, the intensity with CEP=0.5$\pi$ is around five orders higher than that with CEP=0. However, the intensity of the first plateau with CEP=0.5$\pi$ is a little lower than that with CEP=0. Even though the maximum value of $E(t)$ with CEP$=0$ is bigger than that with CEP=$0.5\pi$ as illustrated in Fig. \ref{Fig5}(a), the cutoff energy of the first plateau in the former is less than the latter.

\begin{figure}
\centering\includegraphics[width=9 cm,height=6 cm]{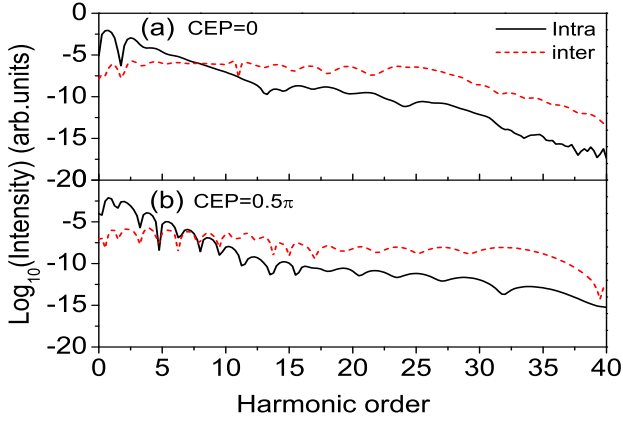}
\caption{(Color online) Interband and intraband current contributions with different CEPs. (a) CEP=0. (b) CEP=$0.5\pi$. Other laser parameters are the same as those in Fig. \ref{Fig5}.} \label{Fig6}
\end{figure}
\begin{figure}
\centering\includegraphics[width=9 cm,height=4.5 cm]{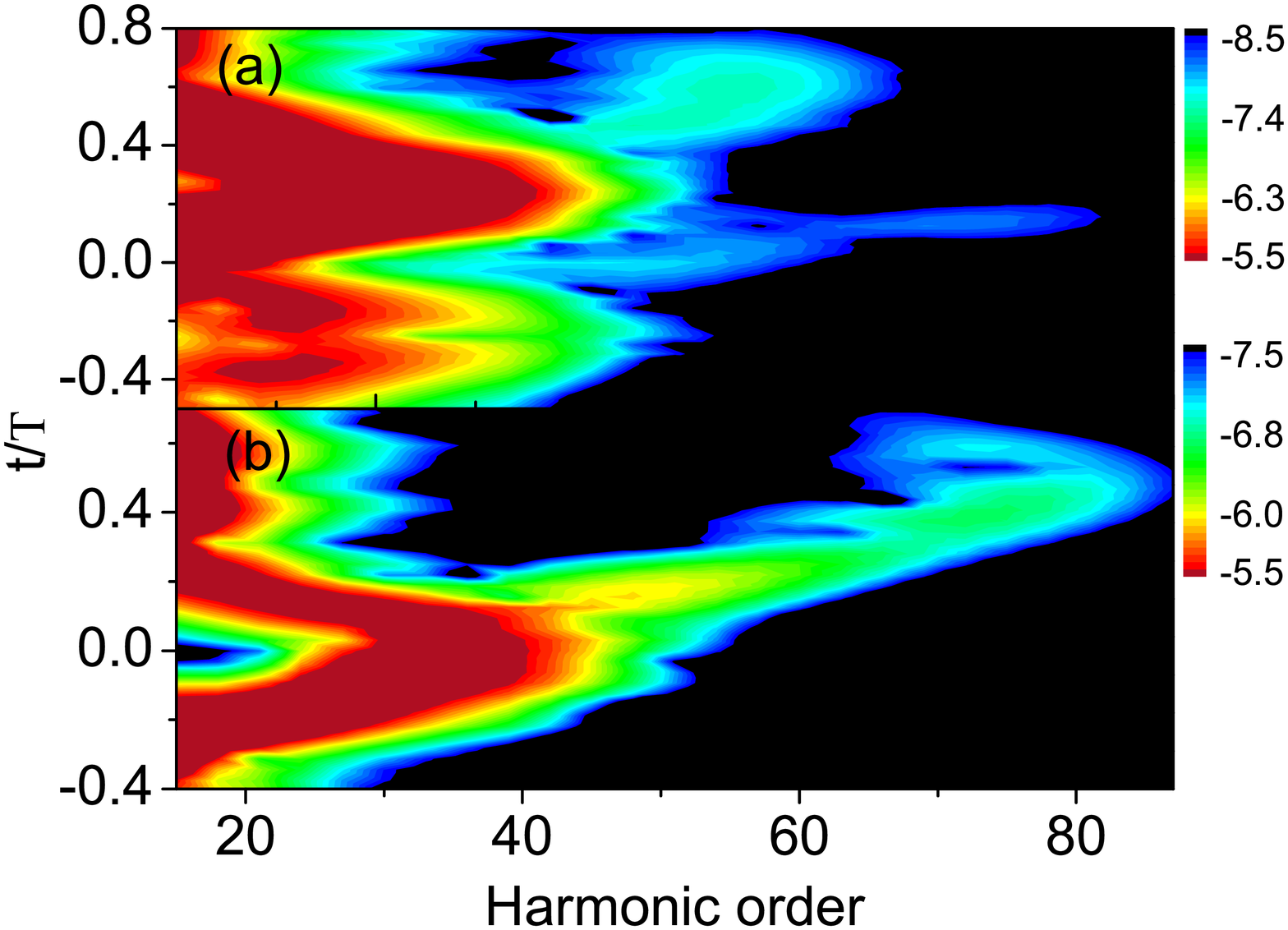}
\centering\includegraphics[width=9 cm,height=4.5 cm]{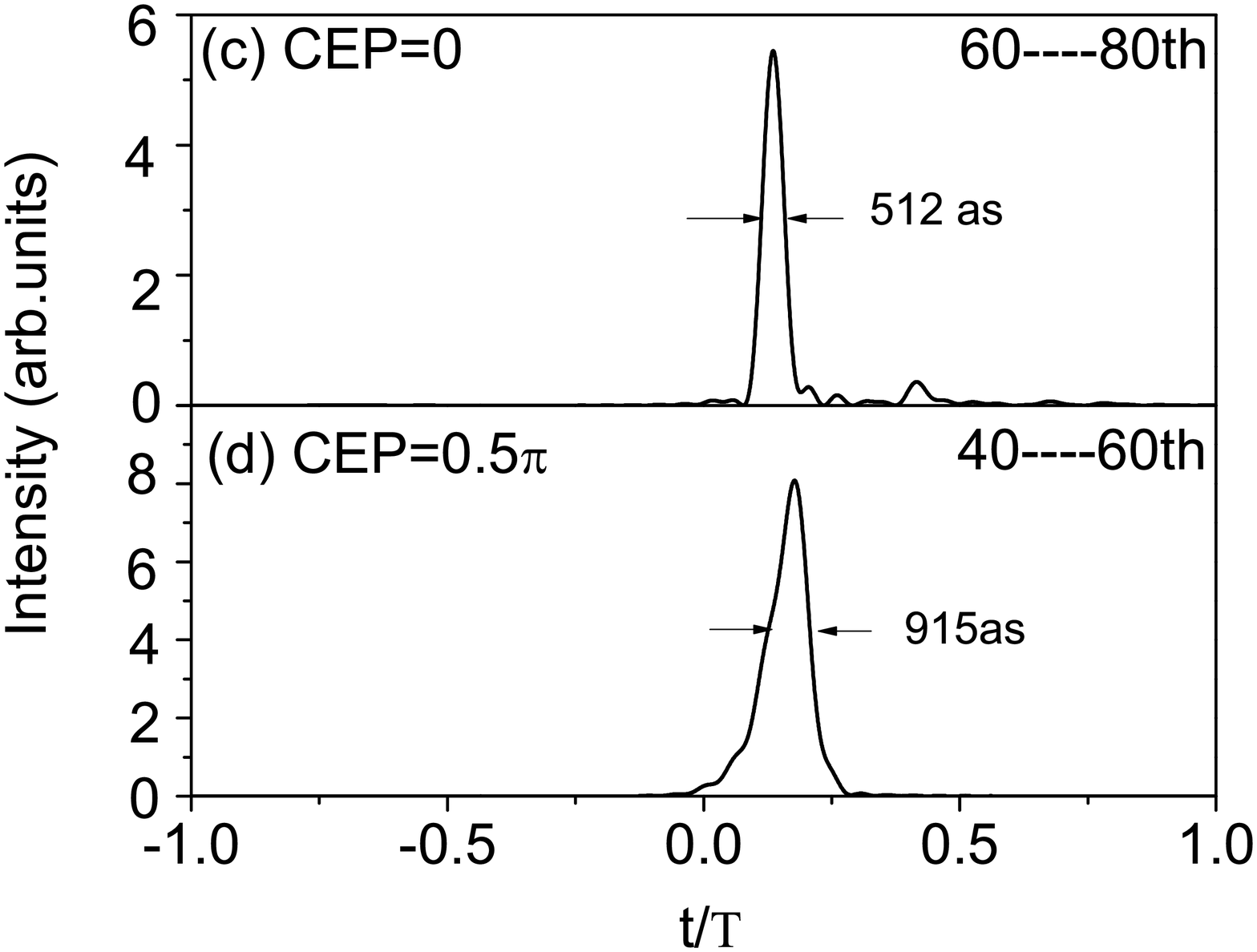}
\caption{(Color online) Time-frequency analysis of HHG in Fig. \ref{Fig5} and isolated attosecond pulse generation. (a) CEP=0. (b) CEP=0.5$\pi$. (c) Attosecond pulse by synthesis of 60-80 HHG with CEP=0. (d) Attosecond pulse by synthesis of 40-60 HHG with CEP=0.5$\pi$.} \label{Fig7}
\end{figure}

In order to further study the physical mechanisms behind these phenomena, we separate interband and intraband current contributions. Since the higher energy bands are not well separated, we only calculate the inter- and intra- contributions in the first plateau. The results are shown in Fig. \ref{Fig6}. One can see that CEP may be used to control the relative contributions from interband and intraband transitions effectively. Next, we perform a time-frequency analysis \cite{Antoine, Uzer} of the two HHG spectra by using the Morlet wavelet \cite{Bian1} and the results are shown in Fig. \ref{Fig7}. From the figures, one can find the "long" and "short" trajectories \cite{Vampa3} as in the case of HHG in gas phase. CEP can be used to control the trajectories as illustrated. This provides us a possible way to generate isolated attosecond pulses from laser-solid interactions. In Figs. \ref{Fig7} (c) and \ref{Fig7} (d), 512 as and 915 as isolated pulses are synthesized with CEP=0 and $0.5\pi$, respectively. Even though the width is much bigger than isolated attosecond pulses generated by HHG in gas phase, it provides us alternative ways to produce intense ultrashort lasers.

\subsection{Laser chirp effect on HHG}
\begin{figure}
\centering\includegraphics[width=9 cm,height=4 cm]{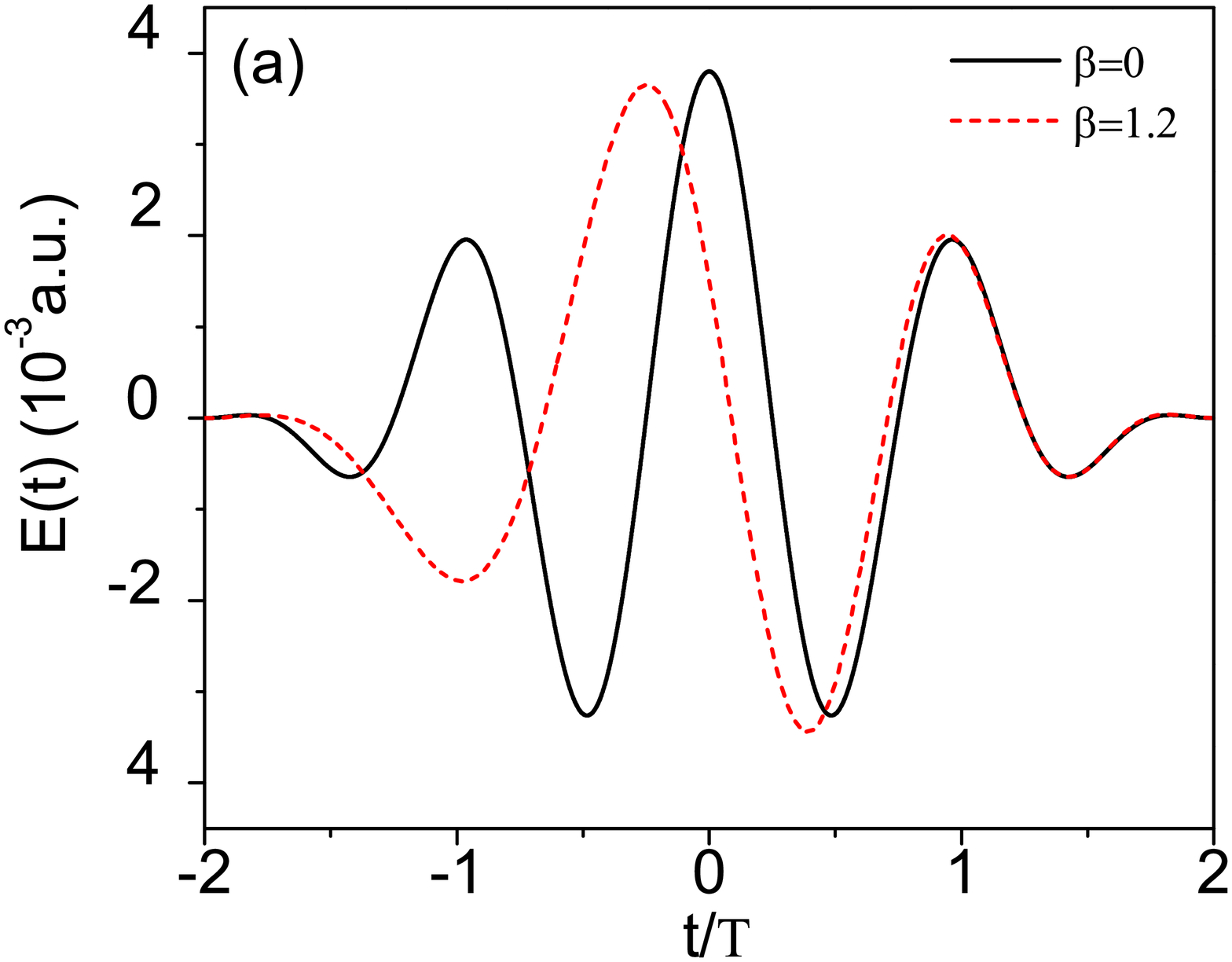}
\centering\includegraphics[width=9 cm,height=4 cm]{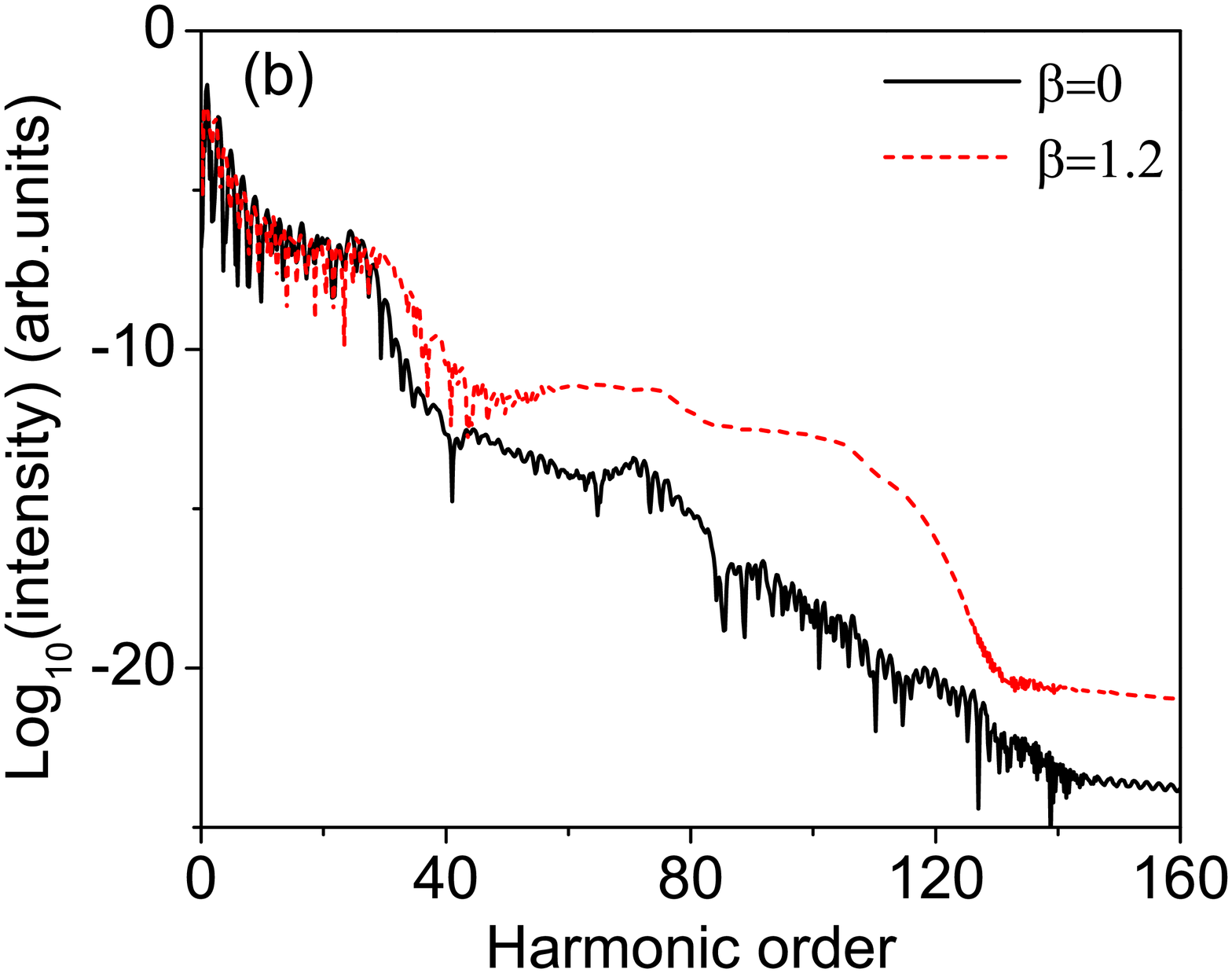}
\centering\includegraphics[width=8 cm,height=5 cm]{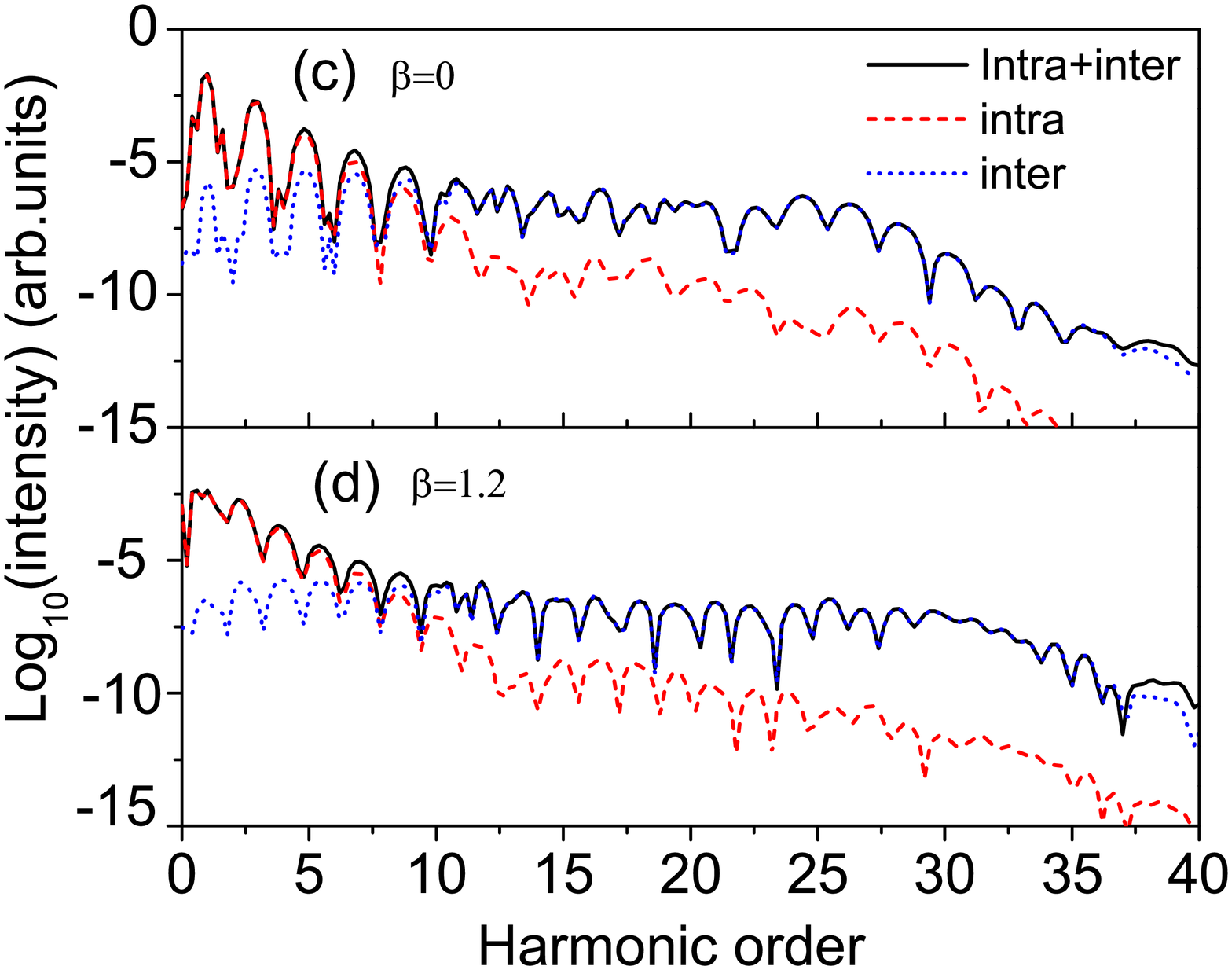}
\caption{(Color online) Laser chirp effect on HHG. (a) Laser fields with different chirps. (b) HHG with different chirps. (c) HHG induced by interband and intraband currents with $\beta=0$. (d) HHG induced by interband and intraband currents with $\beta=1.2$.} \label{Fig8}
\end{figure}

\begin{figure}
\centering\includegraphics[width=9 cm,height= 5 cm]{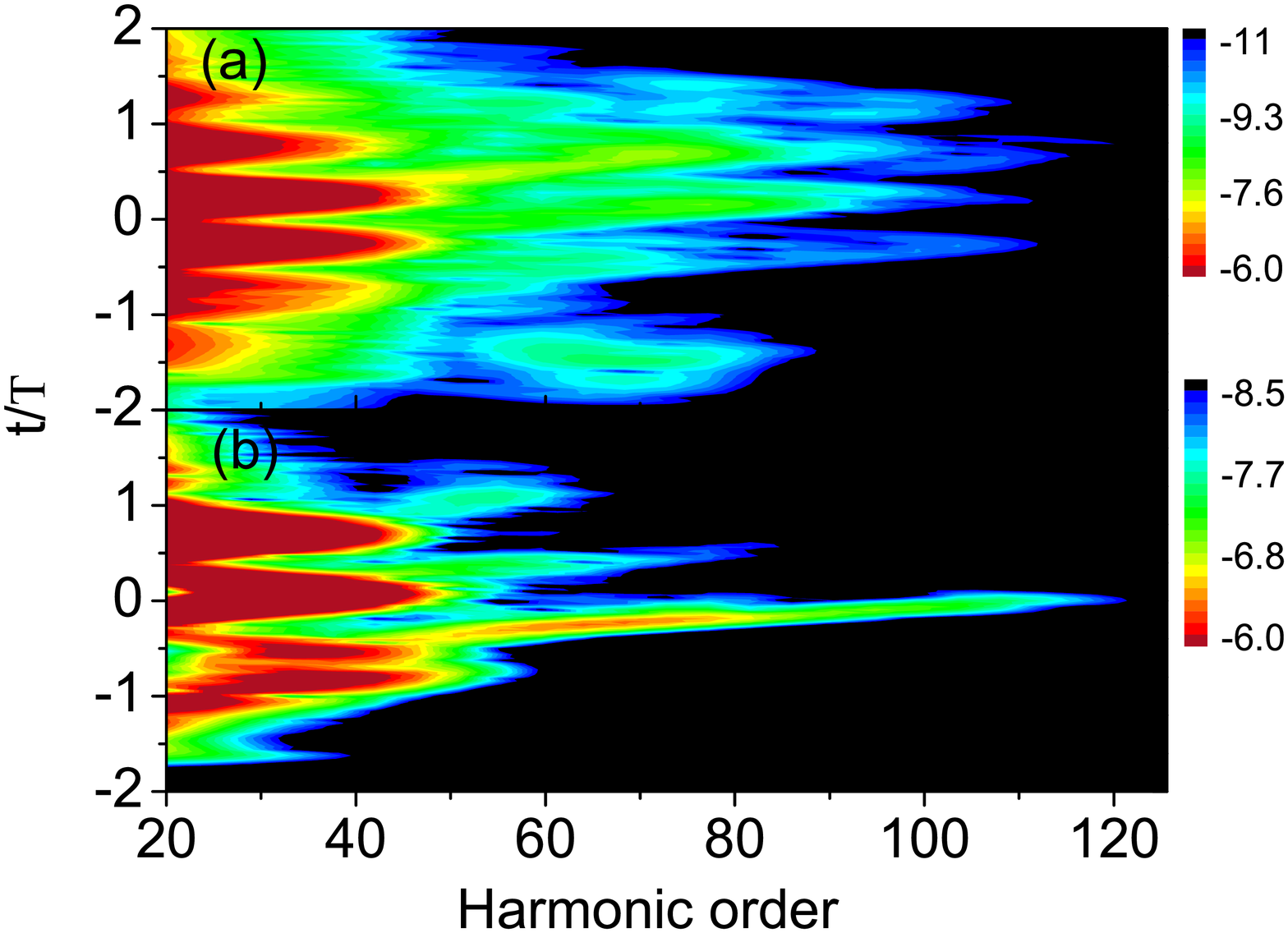}
\centering\includegraphics[width=9 cm,height= 5 cm]{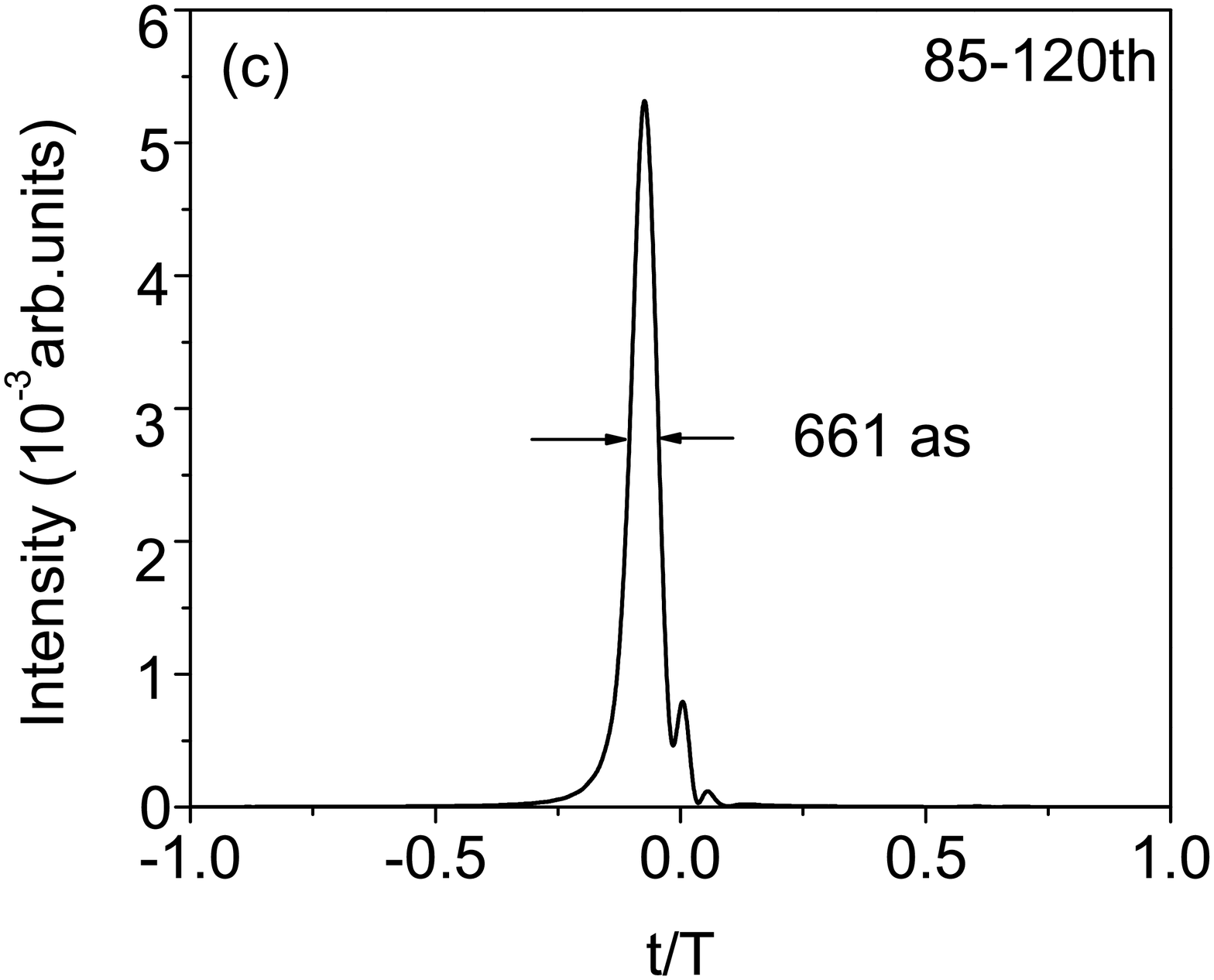}
\caption{(Color online) Time-frequency analysis of HHG in Fig. \ref{Fig8} and isolated attosecond pulse generation. (a) $\beta=0$. (b) $\beta=1.2$. (c) Attosecond pulse by synthesis of 85-120 HHG with $\beta=1.2$.} \label{Fig9}
\end{figure}

In the following we use B-spline method to study laser chirp effect \cite{Peng} on HHG in periodic potentials. In our calculations, the laser field we use is
\begin{equation}
E(t)=E_0f(t)\cos(\omega t+\phi(t)),
\end{equation}
where $\phi(t)=\beta ((t-t_0)/\tau)^2$. The linear chirp is extensively used in experiments \cite{Chang} and calculations \cite{Li}, we choose $t_0$=600 a.u., and $\tau$=610 a.u. The total pulse duration is 4 cycles. In Fig. \ref{Fig8}(a) we show the laser field with different chirps. The HHG spectra with different laser chirp pulses are illustrated in Fig. \ref{Fig8}(b). One may find that the cutoff energy and intensity of the second plateau change dramatically. we also present interband and intraband current contributions to the first plateau in Figs. \ref{Fig8}(c) and \ref{Fig8}(d). It provides us another effective way to control their relative contributions. The time-frequency analysis of the two HHG spectra are shown in Fig. \ref{Fig9}. The chirp effect can be used to control the emission time of HHG and each recombination trajectory. An isolated attosecond pulse with duration 661 as is produced with $\beta=1.2$.

\section{DISSCUSSION}\label{IV}
We simulated the HHG from solid phase based on single-active-electron model in periodic potentials. The corresponding TDSE is solved by B-spline method in coordinate space. The results agree well with Bloch-state expansion in vector $k$ space. It can extract inter and intra band transitions directly, and it is more stable than Houston state expansion. We studied the dynamics of HHG from solid phase in ultra-short few-cycle pulses. To our knowledge, it has not been reported. CEP and laser chirp effects can dramatically change the cutoff energy and intensity of HHG, especially for the second plateau. They can be used to control the relative contributions of inter and intra band transitions, recollision trajectories, and emission times of HHG. The second plateau in solid HHG has not been experimentally identified due to its lower intensity. This work sheds light on how to enhance it. It also shows promising ways to generate isolated attosecond pulses from solid HHG.

\section{ACKNOWLEDGEMENTS}\label{V}
The authors thank Dr. Cheng Gong and Taoyuan Du very much for helpful discussions. This work is supported by the National Natural Science Foundation of China(No. 11404376, No.11465016).

\end{document}